# WHAT DO WE KNOW ABOUT WAVE FUNCTION NODES?

DARIO BRESSANINI

*Dipartimento di Scienze Chimiche, Fisiche e Matematiche - Universita' dell'Insubria, ITALY*
*Georgetown University, Washington, D.C.*

DAVID M. CEPERLEY

*University of Illinois, Urbana-Champaign*

PETER J. REYNOLDS

*Georgetown University, Washington, D.C. and Office of Naval Research*

## 1 Introduction

### 1.1 Nodes and the Sign Problem

Although quantum Monte Carlo is, in principal, an exact method for solving the Schrodinger equation, it is well-known that systems of Fermions still pose a challenge. Thus far all solutions to the *"*sign problem*"* [1] remain inefficient (or wrong). The fixed-node approach [2], however, *is* efficient, and in many situations remains the best approach. If only we could find the exact nodes—or at least a systematic way to improve the nodes—we would, in effect, bypass the sign problem.

### 1.2 The Plan of Attack

A reasonable place to begin is to study the nodes of both exact and good quality approximate trial wave functions. If we can understand their properties, we can perhaps find a way to parameterize the nodes using simple functions. Then one can optimize the nodes by minimizing the fixed-node energy. Unfortunately, very little is known about wave function nodes [3,4], and a systematic study has never been attempted, despite the obvious consequences for improving quantum simulations that such knowledge might generate.

*1.3  The Helium Triplet*

The exact spatial eigenfunctions for the helium atom (and all other two-electron atomic ions) are functions of six coordinates (for two-electron systems the spin eigenfunction can always be factored out). Thus,

$$\Psi_n(\mathbf{R}) = \Psi_n(x_1, y_1, z_1, x_2, y_2, z_2).$$

(Note that bold typeface indicates a vector quantity in any number of dimensions.)

Let us consider states with $S$ symmetry ($L=0$). These states are rotationally invariant, so we can factor out the three Euler angles from the total wave function, and use only three coordinates to describe the internal wave function. These are usually chosen to be the interparticle distances, so that

$$\Psi_n = \Psi_n(r_1, r_2, r_{12}).$$

We now focus our attention on the first triplet state. By the Pauli principle, the wave function must change sign if we interchange the two electrons. It follows that

$$\Psi(r_1, r_2, r_{12}) = -\Psi(r_2, r_1, r_{12}).$$

From this equation we can infer that if we place the two electrons at the same distance from the nucleus (but not necessarily at the same point) we obtain

$$\Psi(r, r, r_{12}) = -\Psi(r, r, r_{12}),$$

implying that

$$\Psi(r, r, r_{12}) = 0.$$

This means that the node is described by the equation $r_1 = r_2$. The first $^3S$ state of helium is one of very few systems where we know the exact node.

Note that the "Pauli hyperplane" $\mathbf{r}_1 = \mathbf{r}_2$, or $\{x_1 = x_2, y_1 = y_2, z_1 = z_2\}$ belongs to the node, but it is only a subset of lower dimensionality. In fact, since we are imposing a single constraint on $\Re^6$ space, the node is a 5-dimensional surface, while the Pauli hyperplane, with three constraints, has dimensionality 3.

Here we summarize some facts about the node of the $2\,^3S$ state, with the main objective to generalize them for larger systems. The big surprise is that the node is more symmetric than the wave function itself, since it does not depend on $r_{12}$. Strikingly, it is also independent of $Z$. Thus He, Li$^+$, Be$^{2+}$,... all have the same node. Furthermore this node is present in all $^3S$ states of two-electron atoms.

While the wave function is not factorizable, i.e.,

$$\Psi(r_1, r_2, r_{12}) \neq f(r_1, r_2) j(r_{12})$$

it *can* be written as

$$\Psi(r_1, r_2, r_{12}) = N(r_1, r_2) e^{f(r_1, r_2, r_{12})}$$

where

$$N(r_1, r_2) = r_1 - r_2 \ .$$

This is not a trivial result. All of the antisymmetry has been placed in a lower dimensional nodal function $N(r_1, r_2)$. The unknown function $f$ is totally symmetric. Moreover, we can write the second factor as an exponential, as we do, to emphasize its positivity. It is also interesting to note that the nodal function $N$ is a simple polynomial in the distances. Last but not least, in this case the Hartree-Fock wave function has the exact node.

### *1.4 Nodal Conjectures*

What properties of the nodes are present in other systems and/or states? In other words, what is general (if anything), and what is specific to atoms versus molecules, to $S$ states versus other symmetries, to triplet states, or to two-electron atoms versus many-electron atoms? Some years ago Anderson [5] found some of these nodal properties in $^1P$ He and $\Sigma_u$ $H_2$ as well. So what is general? For a generic system, what can we say about $N$? Here

$$\Psi_{Exact} = N(\mathbf{R}) e^{f(\mathbf{R})} \ .$$

What we call the "strong nodal conjecture" is the generalization from helium that the exact wave function can be written in the above form, with $N$ an antisymmetric polynomial of finite order, and $f$ a totally symmetric function. In the next sections we will present evidence that the ground states of Li and Be have simple nodes. A weaker conjecture is that $N$ may not be a polynomial, but can be closely approximated by a lower-order antisymmetric polynomial.

### 2   Lithium Atom Ground State

The restricted Hartree-Fock (RHF) description of the ground state of Li is

$$\Psi_{RHF} = \left| 1s(r_1) \overline{1s(r_2)} 2s(r_3) \right| \boldsymbol{aba} = \left( 1s(r_1) 2s(r_3) - 1s(r_3) 2s(r_1) \right) 1s(r_2)$$

As for helium, the node is $r_1 = r_3$, where 1 and 3 are the two *alpha* spin electrons. In other words, if two like-spin electrons are at the same distance from the nucleus then $\Psi = 0$. For triplet He this was an exact result. How good is this RHF node for Li?

As we demonstrate here, even though $\Psi_{RHF}$ is not very good (for example, it belongs to a *higher* symmetry group than the exact wave function), its node is surprisingly accurate. To see this numerically, for example, the exact (DMC) solution with this node gives an energy $E_{RHF} = -7.47803(5)$ a.u. [6] compared to $E_{Exact} = -7.47806032$ a.u. [7]. A DMC simulation (done still with the same nodes), using a Hylleraas function for importance sampling, gives an extrapolated energy $E = -7.478060(3)$ a.u. Thus, the numerical evidence is that the HF node *is* correct (within the error bars of these Monte Carlo simulations). If true, the node has even *higher* symmetry than $\Psi_{RHF}$, since it does not depend on either $\mathbf{r}_2$ or $r_{ij}$.

Surprisingly (perhaps), the GVB wave function has the exact same node. Is this an indication that this node is exact? Not necessarily, since permutational symmetry alone does *not* require this node. The exact wave function, to be a pure $^2S$, requires only that

$$\Psi = f(r_1, r_2, r_3, r_{12}, r_{13}, r_{23}) + f(r_2, r_1, r_3, r_{12}, r_{23}, r_{13}) \\ - f(r_3, r_2, r_1, r_{23}, r_{13}, r_{12}) - f(r_2, r_3, r_1, r_{23}, r_{12}, r_{13})$$

which does not constrain a node at $r_1 = r_3$.

To study an "almost exact" node we took a Hylleraas expansion for Li with 250 terms, whose variational energy $E_{Hy} = -7.478059$ a.u. We then examined the 5-D nodal structure of this function in two ways. First, we used Mathematica® to take cuts through space; and second, we used Monte Carlo simulation to compare nodal crossings of $\Psi_{Hy}$ with crossings of $r_1 = r_3$. From the various cuts, we were unable to find any deviation from the $r_1 = r_3$ node. However from the simulations there were found 6 (out of 98) nodal crossings of the Hylleraas function that did not also cross $r_1 = r_3$. These six crossings appear to be, on closer examination, either sufficiently close to $r_1 = r_3$ to be due to round-off error of a truncated Hylleraas expansion, or to be numerical artifacts. The issue thus appears unresolved by these numerical means.

White and Stillinger [9] determined the nodes for 3 electrons interacting with harmonic springs (so-called *harmonic lithium*) and gave an argument, using perturbation theory, that $r_1 = r_3$ is *not* the exact node for the Li atom ground state. They found that correlations with an electron at $\mathbf{r}_2$ distorted the HF nodes to an aspherical, pear-like shape, but by a small amount. In the vicinity of the nucleus they obtained a form for the nodal function that to lowest order is still a sphere (in $\Re^3$, fixing electrons 2 and 3), but the center is displaced from the nuclear position along the axis in the direction of the third electron, with the radius going through the position of the other up electron.

Given the perturbation results of White and Stillinger, the "strong nodal conjecture" seems not to hold, at least not for the function $r_1 = r_3$ (although it might be true for an higher order polynomial), however the "weak nodal conjecture" is certainly true given the extremely good energy obtained using that node.

## 3  Beryllium Atom

### 3.1  Numerical arguments and Mathematica® cross-sections

The ground state of Be is $(1s)^2 (2s)^2\ ^1S$. In 1992 it was realized that the Hartree-Fock (HF) wave function has *four* nodal regions [3]. In essence, $\Psi_{HF}$ factors into two determinants, each one in effect a triplet $Be^{+2}$. Thus there is (in this description) no effect of the *alpha* electrons on the *beta* electrons, and vice versa, and each set forms a separate $r_1$- $r_2$ type node. Explicitly, the HF node is $(r_1-r_2)(r_3-r_4)$. This node has the *wrong* topology. How do we know this?

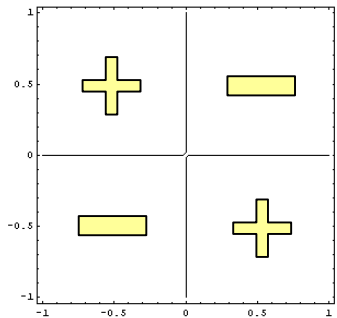

Figure 1: Schematic of nodal regions for $\Psi_{HF}$ of the Be atom. Axes are $t_1=(r_1-r_2)$ and $t_2=(r_3-r_4)$.

Numerically, the DMC energy for this node is $-14.6576(4)$ [6] versus the exact energy of $-14.6673$. This is well outside the DMC statistical error. Since only the fixed node approximation can account for the error, Be represents an unusual system where the nodal error is clearly visible, presumably because of the strong mixing of several HF configurations in the ground state.

In fact, this node is wrong in an easily describable way. It was conjectured a while ago [3,4] that the exact $\Psi$ for ground states of atoms in general have but *two* nodal regions. As indicated, the Hartree-Fock node creates four nodal regions.

We find upon going beyond HF, that the simple node, with its $(r_1-r_2)(r_3-r_4)$ structure, changes only slightly. Yet, the crossing surfaces open up, leaving only a lower dimensional crossing "point." We see this clearly by taking cuts (done in Mathematica®) through the full 9-D (effective) space of the wave function. (The full dimensionality is 3N for an N electron system, and hence 12 here. However, as discussed earlier, for an *S* state the wave function is invariant under rotation, allowing us to eliminate 3 degrees of freedom.)

We started by using Mathematica® to examine gradually more accurate trial wave functions. Plotting $t_1 = (r_1-r_2)$ against $t_2 = (r_3-r_4)$ we see that the HF wave function vanishes along the axes (for arbitrary values of the variables representing the other 7 dimensions). This was illustrated above (though you have to imagine the other 7 dimensions all coming out of the page at you). Given that adjacent regions have opposite signs of the wave function, one can label the regions "+" and "−" as indicated.

An optimized two-configuration (4 determinant) trial wave function also displays this crossing structure, but now only at *particular* values of the other variables. For more general values of those variables there is a passage between either the two "+" regions or the two "−" regions. With a proper choice of the angular variables one observes a smooth opening up of the crossing, going from interconnected "+" regions to interconnected "−" regions.

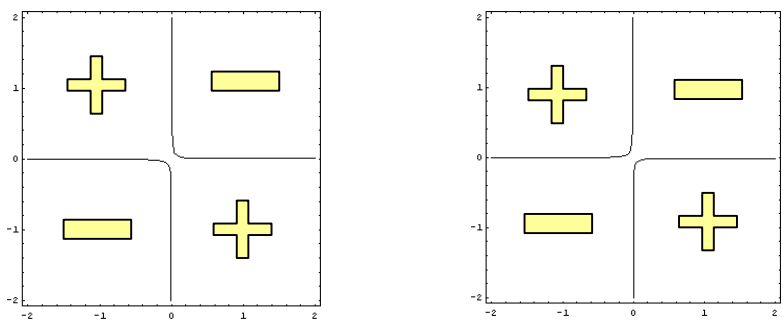

Figure 2: Schematics of nodal regions for $\Psi_{CI}$ of the Be atom. Axes are $t_1=(r_1-r_2)$ and $t_2=(r_3-r_4)$.

The closeness of the nodes to the simple form $(r_1-r_2)(r_3-r_4)$ seems to indicate the presence of this term, plus a small additional term. Taking more cuts provided us with clues. For example there is (almost) a node when two *alpha* electrons are along any ray from the origin, while the two *beta*'s are on any sphere. Following up on this we deduced a node of the form $N(\mathbf{R}) = (r_1 - r_2)(r_3 - r_4) + a\,\mathbf{r}_{12}\cdot\mathbf{r}_{34}$. We are in the process of determining the remaining structure of the node when both these terms are accounted for. A simple polynomial form—of greater symmetry than apparently required (i.e. needing fewer than $d - 1 = 8$ variables)—appears to describe the node.

In the next section we give a proof of what we found numerically; namely that the nodal structure of Be has only two disjoint volume elements. By examining cross-sections it was possible to visualize this, though it remains difficult to understand how the various nodal regions are connected up in the full 12 dimensional space. The proof puts this on firmer footing and also shows the origin of the dot-product term in the node.

### 3.2   *Proof that four-electron $^1S$ atomic ground states have only two nodal regions*

For a singlet state of a four-electron system there will be two up electrons, at positions we denote by $\{\mathbf{r}_1, \mathbf{r}_2\}$, and two down electrons at $\{\mathbf{r}_3, \mathbf{r}_4\}$. The wave function must be antisymmetric with respect to exchange of either of these pairs, with corresponding permutation operators $P_{12}$ and $P_{34}$.

We define a nodal region with respect to a reference point, $\mathbf{R}^* = (\mathbf{r}_1, \mathbf{r}_2, \mathbf{r}_3, \mathbf{r}_4)$, as the set of points that can be reached by a path from the reference point that does not cross a node of $\Psi$. Any point with $\Psi(\mathbf{R}^*) \neq 0$ can be chosen as a reference point. For any ground state wave function, the tiling theorem applies [4]. This theorem tells us that any nodal region defined with respect to one reference point is equivalent to those defined with respect to another point, up to a permutation. For the 4-electron case, this implies that there can be at most four nodal regions, since there are only four permutations that do not interchange spin: two positive permutations $I$, and $P_{12}P_{34}$, and two negative permutations $P_{12}$ and $P_{34}$.

Now, consider the Be atom wave function expanded in a single particle basis. Since the single particle levels are ordered as (1s) < (2s) < (2p) < (3s) …, the two lowest energy configurations in the ground state are $\varphi_1 = (1s)^2(2s)^2$ and $\varphi_2 = (1s)^2(2p)^2$. Explicitly writing down the Slater determinant we find that the sign of $\varphi_1$ is given by the node mentioned above: $(r_1-r_2)(r_3-r_4)$. This gives the four nodal regions of the HF wave function, resulting from the direct product of the up spin and down spin determinants. As stated earlier, this property is not in the true function, wherein electron correlation changes the connectivity of the nodes.

To show that there are, in fact, only two nodal regions we must find a valid reference point $\mathbf{R}^*$ together with a path $\mathbf{R}(t)$ that connects it with its permuted image $P_{12} P_{34} \mathbf{R}^*$ such that $\Psi(\mathbf{R}^*(t)) \neq 0$ along the entire path. (The connection between the two negative regions follows by symmetry.) For the reference point, we consider a point of the form $\mathbf{R}^* = (\mathbf{r}_1, -\mathbf{r}_1, \mathbf{r}_3, -\mathbf{r}_3)$ where $\mathbf{r}_1$ and $\mathbf{r}_3$ are arbitrary non-zero vectors.

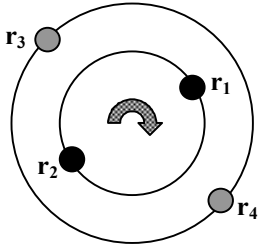

Figure 3: connecting $\mathbf{R}^*$ to $P_{12} P_{34} \mathbf{R}^*$

As an $S$ state, $\Psi(\mathbf{R})$ is invariant with respect to rotation of all of the electrons about any axis through the nucleus. Therefore, consider the path connecting $\mathbf{R}^*$ to $P_{12} P_{34} \mathbf{R}^*$ to be that generated by a $180^o$ rotation about the axis $\mathbf{r}_1 \times \mathbf{r}_3$. (If $\mathbf{r}_1$ is parallel to $\mathbf{r}_3$ then any axis perpendicular to either vector can be chosen.) Since the wave function is invariant with respect to rotation, it is constant along this path. Hence as long as $\mathbf{R}^*$ is a valid reference point, then there are only two connected nodal regions, a single positive and the complementary negative region. This wave function would be *"maximally connected."* What is left in proving that the HF nodes have the incorrect connectivity amounts to finding such a special point with $\Psi(\mathbf{R}^*) \neq 0$. The point $\mathbf{R}^*$ defined above, however, is not suitable if we are to believe the HF node.

We now argue that $\mathbf{R}^*$ from above *is* suitable, as long as electron correlation causes configuration mixing in the ground state. Suppose we expand the exact wave function in a CI basis, $\Psi = \sum_i c_i \boldsymbol{j}_i$. Since the first configuration alone gives just the HF wave function, with $\varphi_1(\mathbf{R}^*) = 0$, we must examine the next term, namely $\boldsymbol{j}_2 = (1s)^2(2p)^2$. We can show that this configuration has a nodal surface described by $(g(r_1)\mathbf{r}_1 - g(r_2)\mathbf{r}_2) \cdot (g(r_3)\mathbf{r}_3 - g(r_4)\mathbf{r}_4) = 0$ for some positive function $g$. With the value of $\mathbf{R}^*$ assumed above, the wave function will vanish only when $\mathbf{r}_1 \cdot \mathbf{r}_3 = 0$. Thus $\mathbf{r}_1$ and $\mathbf{r}_3$ can be freely chosen (as long as they are not perpendicular) for the point $\mathbf{R}^*$ to be a valid reference point. Once electron correlation is included, there will be some amplitude of $\varphi_2$ in $\Psi$, since it has the same symmetry as the ground state, and it is a double excitation. In addition, there is no reason to suspect that adding other terms will cause $\Psi(\mathbf{R}^*)$ to vanish exactly for all of these points, $\mathbf{R}^*$.

The Be atom is a case where the HF single determinant nodes are particularly bad because of configuration mixing, and the HF nodes cause significant fixed-node error [8]. The argument just presented, however, is quite general, applying to all 4-electron atoms having ground state $^1S$ states, and non-zero mixing of double excitations.

## 4   Conclusions

The belief that "nodes are weird" expressed jokingly by M. Foulkes at the Seattle meeting in 1999 may be overstated. Here (at the Hawaii Pacifichem 2000 meeting) we countered with "...maybe not.". Numerically determined nodes (at least for the atoms He, Li, and Be) seem to depend on few variables, and have higher symmetry than the wave function itself. Moreover, the nodes resemble polynomial functions. Possibly this is an explanation of why HF nodes—as seen in fixed-node QMC simulations of these atoms—are so good: they "naturally" have these properties. If so, it is a simple leap of faith to believe that it may in fact be possible to optimize the nodes directly, for use in QMC.

## Draft Index